\documentstyle[preprint,aps]{revtex}
\title{Subthreshold kaon production and the nuclear equation of state}
\author{G. Q. Li and C. M. Ko}
\address{\it{
Cyclotron Institute and Physics Department\\  Texas A\&M University,
College Station, Texas 77843}}
\begin{document}
\maketitle

\begin{abstract}
We reexamine in the relativistic transport model the
dependence of kaon yield on the nuclear equation of state
in heavy ion collisions at energies that are below
the threshold for kaon production from the nucleon-nucleon
interaction in free space.
For Au+Au collisions at 1 GeV/nucleon, we find that the kaon yield
measured by the Kaos collaboration at GSI
can be accounted for if a soft nuclear equation of state
is used.  We also confirm the results obtained in the non-relativistic
transport model that the dependence of kaon yield on the nuclear
equation of state
is more appreciable in heavy ion collisions at lower incident energies.
We further clarify the difference between the predictions from the
relativistic transport model and the non-relativistic transport
model with a momentum-dependent potential.
\end{abstract}

\newpage
About ten years ago, Aichelin and Ko \cite{AICH85}
pointed out that in heavy ion collisions at incident energy per nucleon
that was below the kaon production threshold
in nucleon-nucleon interaction in free space (which is about 1.58 GeV)
kaon production was sensitive to the nuclear equation of state (EOS)
at high densities. In the non-relativistic Boltzmann-Uehling-Uhlenbeck
(BUU) model, they found that
in central heavy ion collisions at an incident energy of 0.7 GeV/nucleon
the kaon yield obtained with a soft EOS (compressibility K=200 MeV)
was about 2-3 times larger than that obtained with
a stiff EOS (K=380 MeV).
This finding was later confirmed in
calculations based on the quantum molecular dynamics
(QMD) \cite{AICH87,LI92}.

The determination of the nuclear EOS at high densities has been one of
the main
motivations for recent experimental measurements of kaons
in heavy ion collisions around 1 GeV/nucleon by the Kaos collaboration
at GSI \cite{GRO93,GRO94}.  Since the experimental data
have become available,
there has been a resurgence of theoretical studies on kaon production in
heavy ion collisions, based on both non-relativistic
\cite{HUANG93,AICH94,LI94} and relativistic \cite{FANG94,MARU94}
transport models.  Huang {\it et al.,} \cite{HUANG93} have carried out
the first comparison of theoretical results, obtained with
the QMD model,
with the experimental data from Au+Au collisions at 1 GeV/nucleon.
Good agreements with the
experimental data have been obtained when a soft EOS is used in
the model. With a stiff EOS, their results are about a factor of two
below the experimental data.
These findings have recently been confirmed by Hartnack {\it et al.,}
\cite{AICH94}, using also the QMD model. A similar calculation has been
carried out by Li \cite{LI94} in the non-relativistic BUU approach. Again,
a soft EOS has been found to give reasonable agreements with
the experimental data.

In both BUU and QMD calculations, the nuclear EOS is modeled by a simple
Skyrme parameterization. In this model, the energy density of the
nuclear matter at density $\rho$ is given by
\begin{eqnarray}
{\cal E}={\alpha \over 2} {\rho ^2\over \rho _0} + {\beta \over \gamma +1}
{\rho ^{\gamma +1}\over \rho ^\gamma _0} + {4\over (2\pi)^3}
\int ^{k_F}_0 d^3{\bbox k}\,(m^2+{\bbox k}^2)^{1/2},
\end{eqnarray}
where $k_F$ is the Fermi momentum and $m$ is the nucleon mass. The parameters
$\alpha , ~\beta $ and $\gamma$ are
determined by requiring that in normal nuclear matter
one has a saturation density of
$\rho _0=0.17$ fm$^{-3}$,
a binding energy of 15.6 MeV, and a compressibility of K=200 MeV (soft
EOS with $\alpha=-356$ MeV, $\beta=303$ MeV, and $\gamma=7/6$) or
380 MeV (stiff EOS with $\alpha=-124$ MeV, $\beta=70.5$ MeV, and
$\gamma=2$).
These two sets of EOS (i.e., the energy per nucleon
as defined by E/A=${\cal E}/\rho-m$) are shown in Fig. 1 by the dashed curves.
At $3\rho _0$, the two differ by about 55 MeV.  We note that
in the Skyrme parameterization, the compressional energy at high densities
depends mainly on the second term in Eq. (1), or more specifically, on
the magnitude of $\gamma$, which is directly proportional to the
nuclear compressibility K at the saturation density.

On the other hand, the calculations of Fang {\it et al.} \cite{FANG94}
and Maruyama {\it et al.} \cite{MARU94} have been
carried out in the relativistic transport model where the nuclear
EOS is modeled using the non-linear $\sigma$-$\omega$ model \cite{QHD}.
The energy density of the nuclear matter in this model is given by
\begin{eqnarray}
{\cal E}&=&{g_\omega ^2\over 2m_\omega ^2} \rho^2+ {m_\sigma ^2\over 2
g_\sigma^2} (m-m^*)^2 +  {b\over 3g_\sigma ^3} (m-m^*)^3
+ {c\over 4g_\sigma ^4} (m-m^*)^4\nonumber\\
&&+{4\over (2\pi)^3}\int ^{k_F}_0 d^3{\bbox k}\,(m^{*2}+{\bbox k}^2)^{1/2},
\end{eqnarray}
where $m^*$ is the nucleon effective mass. The parameters $g_\sigma$,
$g_\omega$, $b$, and $c$ are determined by not only the normal nuclear matter
properties, such as the saturation density, the binding energy,
and the compressibility, but also the nucleon effective mass.
In Ref. \cite{FANG94},
we have used two sets of parameters which correspond to the same nucleon
effective mass, $m^*=~0.83~m$,
but different values of nuclear compressibility, i.e., K=380 MeV
(with $C_\sigma=(g_\sigma/m_\sigma)m=11.27$,
$C_\omega=(g_\omega/m_\omega)m=8.498$, $B=b/(g_\sigma^3m)=-2.83\times
10^{-2}$, and $C=c/g_\sigma^4=0.186$) and
K=200 MeV (with $C_\sigma=13.95$, $C_\omega=8.498$, $B=1.99\times
10^{-2}$, and $C=-2.96\times 10^{-3}$).  It has been found
in Ref. \cite{FANG94} that the kaon yield at 1 GeV/nucleon
obtained with K=200 MeV is only about 15\% larger than that
with K=380 MeV.  This result is inconsistent
with the findings from the non-relativistic transport models
\cite{HUANG93,AICH94,LI94}, where the difference between the results
obtained with the two Skyrme EOS's is about a factor of two.

It has recently been realized that the so-call ``stiff" EOS used in Ref.
\cite{FANG94} is not as stiff as that in the Skyrme parameterization.
As can be seen from Eq. (2), in the relativistic approach
the EOS at high densities largely depends on the vector repulsion
from the omega meson. From the Hugenholtz-Van Hove theorem, which
requires that the Fermi energy is equal to the average single particle
energy at saturation, we have the following relation \cite{GLEN}
\begin{eqnarray}
\Big({g_\omega \over m_\omega} \Big)^2 ={m-E_F^*(\rho _0)-{\bar B}
\over \rho _0},
\end{eqnarray}
where $\rho _0$ and ${\bar B}$ are the saturation density and binding energy,
respectively, and $E_F^*=\big(m^{*2}+k_F^2\big)^{1/2}$.
Since the two EOS's in Ref. \cite{FANG94} have the same nucleon effective
mass, the vector coupling constants are thus the same. At 3$\rho _0$, the
energy per nucleon with the ``stiff" EOS is only about 15 MeV larger than
that of the soft EOS.  This is much smaller than the difference between the
stiff and the soft EOS in the Skyrme parameterization (cf. dashed curves
in Fig.1). As a result, the kaon yield is not very sensitive to the
relativistic nuclear EOS's used in Ref. \cite{FANG94}.

Up to 4$\rho _0$ (see Fig. 1),
the soft EOS used in Ref. \cite{FANG94} is very
close to the soft EOS given by the Skyrme parameterization (Eq. (1)).
To compare results
between relativistic and non-relativistic approaches, we need to
use a stiff EOS in the relativistic approach which is also similar to the
stiff EOS used in the non-relativistic approach.  Such an EOS can be obtained
by using a smaller nucleon effective mass $m^*=~0.68~m$ but the same
nuclear compressibility (K=380 MeV) at saturation density.
The parameters for this EOS are
$$C_\sigma =15.94, ~C_\omega =12.92, ~ B=8.0\times 10^{-4}, ~
C=2.26\times 10^{-3}.$$
The two relativistic EOS's are shown in Fig. 1
by the solid curves. At 3$\rho_0$, the two differ by about 56 MeV as in the
Skyrme parameterization.

To see the sensitivity of subthreshold kaon production
to the two relativistic nuclear EOS's, we have carried out a
perturbative calculation of kaon production in Au+Au collisions
using the relativistic
transport model developed in Ref. \cite{KO87}.
Kaons are mainly produced from
baryon-baryon interactions, and the production cross sections are
taken from the linear parameterization
of Randrup and Ko \cite{KO80}.
Contributions from meson-baryon interactions \cite{XIONG}, higher
resonances \cite{BALI}, and multi-baryon interactions \cite{BATKO}
have been neglected as they are unimportant for kaon production at energies
around 1 GeV/nucleon.
The rescattering of produced kaons with nucleons is treated by the
perturbative test particle method introduced in \cite{ZHENG}.
Details of the calculations can be found in Ref. \cite{FANG94}.

For a head-on Au+Au collision at
1 GeV/nucleon, we show in the left panel of Fig. 2 the total number of
baryon-baryon collisions that have energies above the kaon
production threshold. With the soft EOS, this number is about 95
but is reduced to about 54 when the stiff EOS is used. The reduction is
partly due to the fact that the maximum central density reached
with the soft EOS (about 2.9$\rho _0$) is higher than that with
the stiff EOS (about 2.4$\rho_0$). As a result,
the average density at which kaons are produced is also higher
for the soft EOS (about 2.5$\rho_0$) than for the stiff one (about
2.1$\rho_0$).  Furthermore, the energy per nucleon at these densities
is about 5 MeV for the former and 15 MeV
for the latter. Thus, more kinetic energy is converted into the
compressional energy in the case of the stiff EOS.
This effect can be seen from the
the right panel in Fig. 2, where we show the distribution of
$p_{max}$ in the collision, with $p_{max}$ being the maximum
momentum of the produced kaon in a given baryon-baryon collision.
The average value of $p_{max}$ is about 0.272 GeV/c in the
case of the soft EOS and is reduced to about 0.245 GeV/c for the stiff EOS.
Overall, the kaon yield with the soft EOS
is about a factor of two larger than that with the stiff one,
consistent with the findings of non-relativistic transport models
\cite{HUANG93,AICH94,LI94}.

Our results thus demonstrate that the kaon yield from heavy ion collisions
is similar in both relativistic and non-relativistic transport models
if similar nuclear equations of state are used.
In Ref. \cite{AICH87}, the non-relativistic transport model was generalized
to include a momentum-dependent potential, and it was shown that this
would reduce significantly the kaon yield due to the lower number of collisions
and deceleration as a result of the momentum-dependent potential.
This result is in contrary to ours
based on the relativistic transport model, which includes
the momentum-dependent potential via the nucleon effective mass.
We believe that the kaon yield calculated in Ref. \cite{AICH87}
is incorrect as it has not taken into account
the difference in the initial and final potential energies in the reaction
$BB\to NYK$, where $B$ and $Y$ denote a baryon (nucleon or delta)
and a hyperon (lambda or sigma), respectively. Since
baryons have larger momenta in the initial state than in the final state,
some of the initial potential energy is available for kaon production and
should compensate for the reduction of kaon yield
due to the momentum-dependent potential.
In our relativistic transport model \cite{FANG94},
these effects are properly treated by including not only the nucleon
effective mass but also the hyperon and kaon effective masses. The reduction
of energy in the initial state of the reaction $BB\to NYK$ due to
reduced nucleon in-medium mass is thus compensated by a corresponding
reduction in the threshold as the final state energy is also reduced
when in-medium masses are used.
The net effect of modified hadron in-medium masses in the reaction is thus
small, and our results are therefore similar to that from the normal
non-relativistic transport model.
However, this does not mean that it is correct to use the normal
non-relativistic transport model to describe kaon production as we know
that the nucleon mean-field potential is momentum-dependent and
should be included. Furthermore, it is incorrect to assume,
as in the non-relativistic transport model, that the
hyperon has the same potential as the nucleon.
{}From the phenomenology of hypernuclei, it is known that the mean-field
potential for the hyperon is only about 2/3 of the nucleon potential
\cite{MIL88,YAM88}. Also, the neglect of kaon potential in the
non-relativistic transport model is not warranted \cite{KAP86}.
It has been recently shown that the kaon potential has significant effects
on the flow of kaons in heavy ion collisions \cite{GQL94}.

In Fig. 3  we compare the kaon
momentum spectra obtained with the soft EOS (the same as reported in Ref.
\cite{FANG94}) and the stiff EOS with the experimental data from the
Kaos collaboration \cite{GRO93,GRO94}. The theoretical results obtained
with the soft EOS are in reasonable agreements with the experimental data,
while that with the stiff EOS are below the data by about a factor
of two. Thus, the Kaos data from Au+Au collisions at 1 GeV/nucleon
favor a soft EOS.  Because of the lack of empirical information on
the elementary kaon production cross section from
the nucleon-nucleon interaction near the threshold,
this conclusion should be taken with some cautions \cite{AICH94}.

As pointed out in Ref. \cite{AICH85}, the difference between the kaon
yields obtained with the soft and stiff EOS's
in the Skyrme parameterization increases with decreasing
incident energy. To see whether this is also the case in a relativistic model,
we have carried out calculations for head-on
Au+Au collisions from 0.6 GeV/nucleon
to 1.2 GeV/nucleon. The results are shown in Fig. 4. The left panel gives the
kaon production probability P$_{{\rm K}^+}$
as a function of the incident energy $E_{\rm inc.}$, and the right panel
gives the ratio between the production probabilities
obtained with the soft and stiff EOS's.
We find that as the incident energy decreases from 1.2 GeV/nucleon to
0.6 GeV/nucleon, the ratio increases from 1.9 to 4.2, similar to the
results from non-relativistic transport models \cite{AICH85}.
The effect of the
nuclear EOS on the kaon yield can thus be more clearly studied at
lower incident energies.

In summary, using two relativistic
EOS's that are similar to the Skyrme-type EOS's used in non-relativistic
transport models, we have shown that in heavy ion
collisions at subthreshold energies
kaon production is sensitive to the nuclear EOS at
high densities. Recent kaon data at 1 GeV/nucleon from the Kaos
collaboration seem to favor a soft EOS.  To learn more definitively about
the nuclear equation of state at high densities, heavy ion experiments at
lower incident energies, such as around 0.6 GeV/nucleon,
will be very useful as kaon production
at these energies is more sensitive to the nuclear EOS.

\medskip

We are grateful to J. Aichelin, G. E. Brown, and V. Koch for helpful
discussions.
This work was supported in part by NSF Grant No. PHY-9212209
and Welch Foundation Grant No. A-1110.
We also thank the [Department of Energy's] Institute for Nuclear Theory
at the University of Washington for its hospitality and the Department
of Energy for partial support during the completion of this work.

\newpage
\medskip\medskip
\centerline{\bf Figure Captions}

\begin{description}
\item{Fig. 1:} ~Nuclear equation of state in the non-linear $\sigma$-$\omega $
model (solid curves) and in the Skyrme parameterization (dashed curves).

\item{Fig. 2:} ~Time evolution of the total number of baryon-baryon collisions
that have energies above the kaon production threshold (left panel),
and the distribution of the kaon maximum momentum $p_{max}$ (right panel).

\item{Fig. 3:} Kaon momentum spectra obtained with the soft EOS (solid curve)
and the stiff EOS( dashed curve). Experimental
data from Ref. \cite{GRO94} are shown with open squares.

\item{Fig. 4:} Kaon production probability as a function of incident energy
(left panel), and the ratio between the kaon production probabilities
obtained with the soft and stiff EOS's.

\end{description}


\bigskip\bigskip

\begin{thebibliography}{99}

\bibitem{AICH85} J. Aichelin and C. M. Ko, Phys. Rev.
Lett. {\bf 55}, 2661 (1985).

\bibitem{AICH87} J. Aichelin, A. Rosenhauer, G. Peilert, H. St\"ocker,
and W. Greiner, Phys. Rev. Lett. {\bf 58}, 1926 (1987).

\bibitem{LI92} G. Q. Li, S. W. Huang, T. Maruyama, Y. Lotfy,
D. T. Khoa, and A. Faessler, Nucl. Phys.  {\bf A537},
645 (1992).

\bibitem{GRO93}E. Grosse, Prog. Part. Nucl.
Phys. {\bf 30}, 89 (1993).

\bibitem{GRO94} D. Miskowiec, {\it et al.,} Phys. Rev. Lett. {\bf 72}, 3650
(1994).

\bibitem{HUANG93} S. W. Huang, A. Faessler, G. Q. Li, R. K. Puri,
E. Lehmann, M. A. Martin, and D. T. Khoa, Phys. Lett.
B {\bf 298}, 41 (1993).

\bibitem{AICH94} G. Hartnack, J. Janicke, and J. Aichelin, Preprint
LPN-93-11, Nucl. Phys. A, in press.

\bibitem{LI94} B.-A. Li, Phys. Rev. C {\bf 50}, 2144 (1994).

\bibitem{FANG94} X. S. Fang, C. M. Ko, G. Q. Li, and Y. M. Zheng,
Phys. Rev. C {\bf 49}, R608 (1994); Nucl. Phys. {\bf A575}, 766 (1994).

\bibitem{MARU94} T. Maruyama, W. Cassing, U. Mosel, S. Teis, and K. Weber,
Nucl. Phys. {\bf A573}, 653 (1994).

\bibitem{QHD} B. D. Serot and J. D. Walecka,
Adv.  Nucl. Phys. {\bf 16}, 1 (1986).

\bibitem{GLEN} N. K. Glendenning, Nucl. Phys. {\bf A512}, 737 (1990).

\bibitem{KO87} C. M. Ko, Q. Li, and R. Wang, Phys. Rev. Lett. {\bf 59}, 1084
(1987);  C. M. Ko and Q. Li, Phys. Rev. C {\bf 37}, 2270 (1988);
Q. Li, J. Q. Wu, and C. M. Ko, Phys. Rev. C {\bf 39}, 849 (1989);
C. M. Ko, Nucl. Phys. {\bf A495}, 321c (1989).

\bibitem{KO80} J. Randrup and C. M. Ko, Nucl. Phys. {\bf A343}, 519 (1980);
{\bf A411}, 537 (1983).

\bibitem{XIONG} L. Xiong, C. M. Ko, and J. Q. Wu, Phys. Rev. C {\bf 42},
2231 (1990).

\bibitem{BALI} B. A. Li, C. M. Ko, and G. Q. Li, Phys. Rev. C, in press.

\bibitem{BATKO} G. Batko, J. Randrup, and T. Vetter, Nucl. Phys. {\bf A536},
786 (1992).

\bibitem{ZHENG} X. S. Fang, C. M. Ko, and Y. M. Zheng, Nucl. Phys.
{\bf A566}, 499 (1993).

\bibitem{MIL88} D. J. Millener, C. B. Dover, and A. Gal, Phys. Rev.
C {\bf 38}, 2700 (1988).

\bibitem{YAM88} Y. Yamamoto, H. Bando, and J. Zofka, Prog. Theor. Phys.
{\bf 80}, 757 (1988).

\bibitem{KAP86} D. B. Kaplan and A. Nelson, Phys. Lett. B {\bf 175}, 57
(1986); A. Nelson and D. B. Kaplan, {\it ibid}, B {\bf 192}, 193 (1987).

\bibitem{GQL94} G. Q. Li, C. M. Ko, and B. A. Li, Phys. Rev. Lett., in press.

\end{thebibliography}
\end{document}